\documentclass{aa}  

\usepackage{graphicx}
\usepackage{natbib}
\usepackage{color}
\usepackage{upgreek}
\usepackage{soul}

\bibpunct{(}{)}{;}{a}{}{,}
\newcommand{\ms}{m\,s$^{-1}$}

\usepackage{txfonts}
\usepackage{mathrsfs}
\begin{document}

\titlerunning{$\upgamma$\,Equ}
\authorrunning{J\"arvinen et al.}

\title{Testing pulsation diagnostics in the rapidly oscillating
  magnetic Ap star $\upgamma$\,Equ using near-infrared CRIRES$+$ observations}

   \author{
          S.~P.~J\"arvinen
          \inst{1}
          \and
          S.~Hubrig\inst{1}
          \and
          B.~Wolff\inst{2}
          \and
          D.~W.~Kurtz\inst{3,4}
           \and
          G.~Mathys\inst{5}
          \and
          S.~D.~Chojnowski\inst{6}
          \and
          M.~Sch\"oller\inst{2}
          \and 
          I.~Ilyin\inst{1}
          }

   \institute{Leibniz-Institut f\"ur Astrophysik Potsdam (AIP),
              An der Sternwarte~16, 14482~Potsdam, Germany\\
              \email{sjarvinen@aip.de}
         \and
             European Southern Observatory,
             Karl-Schwarzschild-Str.~2, 85748 Garching, Germany
         \and
             Centre for Space Physics, North-West University, 
             Mahikeng 2735, South Africa
         \and
            Jeremiah Horrocks Institute, University of Central Lancashire,
             Preston PR1 2HE, UK
          \and
             European Southern Observatory,
             Alonso de Cordova 3107, Vitacura, Santiago, Chile   
           \and NASA Ames Research Center, Moffett Field, CA 94035, USA
}

   \date{Received 30 December 2023 / Accepted 16 January 2024}

 
  \abstract
{
Pulsations of rapidly oscillating Ap stars and their interaction with the 
stellar magnetic field have not been studied in the near-infrared (near-IR)
region despite the benefits these observations offer compared to visual 
wavelengths. The main advantage of the near-IR is the quadratic dependence 
of the Zeeman effect on the wavelength, as opposed to the linear dependence 
of the Doppler effect.
   }
   {
To test pulsation diagnostics of roAp stars in the near-IR, we aim to 
investigate the pulsation behaviour of one of the brightest magnetic roAp 
stars, $\upgamma$\,Equ, which possesses a strong surface magnetic field of 
the order of several kilogauss and exhibits magnetically split spectral 
lines in its spectra.
}
   {
Two magnetically split spectral lines belonging to different elements, the 
triplet \ion{Fe}{i} at 1563.63\,nm and the pseudo-doublet \ion{Ce}{iii} at 
1629.2\,nm, were recorded with CRIRES$+$ over about one hour in the 
$H$ band with the aim of understanding the character of the line profile 
variability and the pulsation behaviour of the magnetic field modulus.
}
   {
The profile shapes of both studied magnetically split spectral lines vary in 
a rather complex manner probably due to a significant decrease in the strength 
of the longitudinal field component and an increase in the strength of the 
transverse field components over the last decade. A mean magnetic field 
modulus of 3.9\,kG was determined for the Zeeman triplet \ion{Fe}{i} at 
1563.63\,nm, whereas for the pseudo-doublet \ion{Ce}{iii} at 1629.2\,nm we 
observe a much lower value of only about 2.9\,kG. For comparison, a mean 
field modulus of 3.4\,kG was determined using the Zeeman doublet 
\ion{Fe}{ii} at 6249.25 in optical PEPSI spectra recorded just about two 
weeks before the CRIRES$+$ observations. Different effects that may lead to 
the differences in the field modulus values are discussed. Our measurements 
of the mean magnetic field modulus using the line profiles recorded in 
different pulsational phase bins suggest a field modulus variability of 
32\,G for the Zeeman triplet \ion{Fe}{i} at 1563.63\,nm and 102\,G for the 
pseudo-doublet \ion{Ce}{iii} at 1629.2\,nm. }
   {}


   \maketitle
%

\section{Introduction}\label{sec:introduction}

The rapidly oscillating Ap (roAp) main-sequence stars belong to a sub-group of
H-core-burning SrCrEu peculiar A stars with $T_{\rm eff}$ in the range of 
about 6600\,K to 8500\,K. They are usually strongly magnetic and pulsate in 
high-overtone, low-degree, non-radial p-modes with periods in the range of 
about 5 to 24 min. They have been observed intensively photometrically since 
their discovery by 
\citet{1978IBVS.1436....1K}, 
who detected a 12 min pulsation period in Przybylski’s star 
(\object{HD\,101065})
using ground-based photometric observations. Frequency analyses of the light 
curves of the roAp stars have yielded rich asteroseismic information on the 
degrees of the pulsation modes, the distortion of the modes from normal 
modes, the magnetic geometries, and the interaction of the pulsation with the 
magnetic field. Whereas several thousands of Ap stars are currently known
\citep[e.g.][]{2009A&A...498..961R},
the number of identified roAp stars is just over 100
\citep[e.g.][]{2015MNRAS.452.3334S,2021MNRAS.506.1073H,Daniel2024}. 

The main property of roAp stars is that they are oblique pulsators, and the 
strength and geometry of their global magnetic fields constrain the 
pulsation modes. 
\citet{1982MNRAS.200..807K}
suggested that the pulsation mode axis is aligned with the magnetic axis, 
which is itself inclined to the rotation axis so that the observer sees the 
pulsation modes from an aspect that varies with rotation. The roAp stars 
show not only photometric variability, but also rapid radial velocity (RV) 
variations with the same mode frequencies as those obtained from photometry. 
The detected pulsation amplitudes of radial velocities depend on the element 
used: the lines of rare earth elements (REEs) and the H$\alpha$ core show 
the highest amplitudes, whereas iron-peak elements with the higher formation 
depth in the atmosphere usually show low RV variations
\citep[e.g.][and references therein]{2003MNRAS.343L...5K}.
This is an indication that the pulsation amplitudes are a function of 
atmospheric depth.

Assuming that the pulsations of roAp stars are governed by their magnetic
fields, several authors obtained spectral time series in circular polarised 
light in a visual spectral region to study magnetic field variations over the 
pulsation cycle in a number of stars. 
\citet{2004A&A...415..661H}
measured the magnetic field variability over the pulsation cycles in six 
roAp stars using low-resolution FOcal Reducer low dispersion Spectrograph
\citep[FORS\,1;][]{1998Msngr..94....1A}
spectral time series; however, only one roAp star, Przybylski's star 
(HD\,101065), showed a signal for magnetic variability with a frequency of 
1.365\,mHz and an amplitude of $39\pm12$\,G.
\citet{2003A&A...407L..67L}
used high-resolution circular polarisation observations of individual lines
in the spectra of the ultra-slowly rotating roAp star $\upgamma$\,Equ 
(\object{HD\,201601}) and reported a clear detection of the mean 
longitudinal field
variability of up to $240\pm37$\,G in the \ion{Nd}{iii} line at 
5845.07\,\AA{} over the pulsation cycle. This discovery was, however,
questioned by
\citet{2004A&A...415L..13K},
who obtained an upper limit of 40--60\,G using 13 \ion{Nd}{iii} lines.

\citet{2005EAS....17..113M,2007MNRAS.380..181M}
studied the pulsational behaviour of the individual $\uppi$ and $\upsigma$
components using the \ion{Eu}{ii} line at 6437\,\AA{} in the spectroscopic
time series of the strongly magnetic roAp star \object{HD\,166473}, which 
exhibits radial velocity variations due to pulsation with three 
frequencies: 1.833, 1.886, and 1.928\,mHz
\citep{2003MNRAS.343L...5K}. 
The results of these studies hinted at the occurrence of variations of the 
mean magnetic field modulus with the pulsation frequency 1.928\,mHz and an 
amplitude of $21\pm5$\,G.

Importantly, more recent studies of the oblique pulsator model showed that 
the mode geometries in roAp stars are complex and should be considered as a 
function of atmospheric height. Using high-resolution time-series spectroscopy,
it became possible to identify the pulsation geometry not only over the 
stellar surface but also vertically in the  stellar atmosphere. As an 
example, high-speed spectroscopy of the roAp star \object{HD\,99563} to 
study the pulsation amplitude and phase behaviour of elements in its 
stratified atmosphere over the 2.91-d rotation cycle was carried out by 
\citet{2009MNRAS.396..325F}.
The authors identified spectral features related to patches in the surface
distribution of chemical elements and studied the pulsation amplitudes and 
phases as the patches moved across the stellar disc. The detected variability 
had been consistent with a distorted non-radial-dipole pulsation mode. 

\citet{2021MNRAS.508L..17H}
took a step further in the analysis of pulsations of roAp stars and 
investigated the usefulness of linear polarisation as an enhanced pulsation 
diagnostic. While the circular polarisation depends upon the magnetic vector 
projection on to the line of sight, the linear polarisation depends upon the 
magnetic vector projected on to the plane perpendicular to the line of sight.
The authors reported on a possible detection of pulsational variability of 
the transversal field component in Fe, Nd, and Eu lines. The detected 
variability had been considered as due to the impact of pulsations on the 
transverse magnetic field, causing changes in the obliquity angles of 
magnetic force lines.

To better understand the interaction between the pulsation and the stellar 
magnetic field and to test the diagnostic potential of near-IR 
observations, we recently carried out high-resolution spectroscopical time 
series of one of the brightest roAp stars, $\upgamma$\,Equ ($m_{v} =4.7$), 
over about one hour in the $H$ band using the upgraded CRyogenic InfraRed 
Echelle Spectrograph
\citep[CRIRES$+$;][]{2023A&A...671A..24D}.
As of today, pulsations of roAp stars and their interaction with the stellar 
magnetic field have not been studied in the near-IR region despite the numerous
advantages such observations offer compared to visual wavelengths. The main
advantage of the near-IR is the quadratic wavelength dependence of the 
Zeeman effect, as opposed to the linear dependence of the Doppler effect. 
This opens the possibility, at a given magnetic field strength, of observing 
resolved lines in faster rotating stars
\citep[e.g.][]{2019ApJ...873L...5C}
or, conversely, at a given projected equatorial velocity, of resolving Zeeman 
split comments in more weakly magnetic stars. Furthermore, performing a 
near-IR study permits the exploration of a different range of photospheric 
depths, thus providing more insight into the depth dependence of pulsation. As 
the detection of the variation of the magnetic field modulus over the 
pulsation period is yet to be confirmed by independent observations, 
high-resolution spectroscopic time series obtained in the near-IR allowing 
the analysis of the pulsational behaviour of the individual $\uppi$ and 
$\upsigma$ components is especially important.

$\upgamma$\,Equ is distinguished among the roAp stars by very sharp lines 
caused by the extremely slow rotation, a rather strong mean longitudinal 
magnetic field varying from about 630\,G to $-1180$\,G
\citep[e.g.][]{2018AstBu..73..463S},
and a magnetic field modulus of 4\,kG measured using magnetically split 
spectral lines resolved in the visible
\citep[e.g.][]{2017A&A...601A..14M}.
The most recent estimate of the rotation period for this star gives a lower
limit of 95\,yr
\citep{2016MNRAS.455.2567B,2018AstBu..73..463S}.
Furthermore, $\upgamma$\,Equ exhibits a rather high amplitude of pulsational 
line profile radial velocities in the visual exceeding 1000\,\ms{} in 
individual REE spectral lines and has well-studied pulsation characteristics,
with the highest amplitude frequency corresponding to a period of about 12 
minutes.

In the following we discuss the available observational material and the 
analysis of the pulsational behaviour of the $\uppi$ and $\upsigma$ components 
observed in the selected magnetically split spectral lines in the 
high-resolution spectroscopic  CRIRES$+$ time series, and we discuss the 
potential usefulness of multi-wavelength observations for the improvement of 
our understanding of the pulsational properties of roAp stars.


\section{Spectroscopic and photometric observations}\label{sec:obs}

To study the pulsation and its interaction with the magnetic field in the 
roAp star $\upgamma$\,Equ, we obtained, on 2022 September 28, high-resolution 
spectroscopic CRIRES$+$ time series over 56\,min in the $H$ band
(from 1438 to 1765\,nm), sampling several lines belonging to iron-peak and 
rare-earth elements. The line identification in the $H$ band was previously 
carried out by  
\citet{2019ApJ...873L...5C}.
Using the nodding cycle ABBA, we obtained 35 spectra with the narrowest 
available slit of 0.2\,\arcsec,{} resulting in a spectral resolution of about 
100\,000. Within the ABBA nodding cadence, the first exposure is recorded 
at the position A; then, it is followed by two exposures at the position B, 
and the fourth exposure is obtained moving back to the position 
A\footnote{\url{https://www.eso.org/sci/facilities/paranal/instruments/crires/doc/CRIRES_User_Manual_P113.1.pdf}}. 
The purpose of nodding -moving the telescope to the positions A and B along the 
direction of the slit- is to remove sky emission, detector dark current and 
glow, and some ghosts. With the exposure time of individual frames of 10\,s 
between the nodding and the full individual sequence, ABBA accounted for about 
1.5\,min. The data were reduced using the CR2RES pipeline recipes. As 
both ABBA and AB/BA nodding cycles are available, the reduction was carried 
out for both cycles. For the AB/BA nodding cycles, we obtained 70 spectra 
with the individual AB/BA cycle length of about 50\,s. The typical 
signal-to-noise ratio ($S/N$) for the spectra obtained with the ABBA cycle 
is 212, whereas the $S/N$ for the individual AB/BA cycles is 153.

\begin{figure}
    \centering
    \includegraphics[width=0.90\columnwidth]{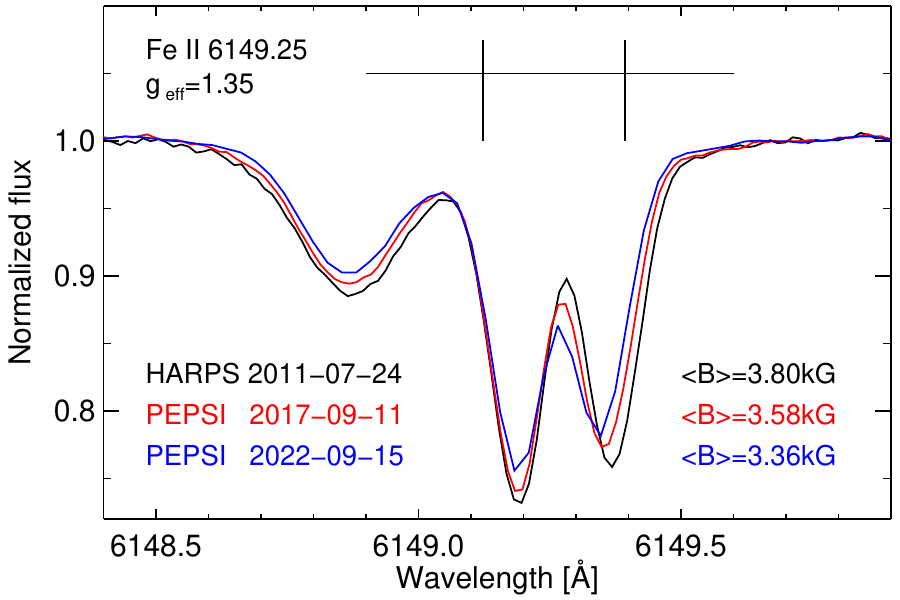}
    \caption{
Zeeman doublet \ion{Fe}{ii} at 6149.25\,\AA{} with a Land\'e factor 
$g_{\rm eff} =1.35$ used for measuring the mean magnetic field modulus in 
observations obtained in different years using different instruments. The 
corresponding Zeeman pattern is displayed above the line profiles, with the 
$\uppi$ components shown above the horizontal line and the $\upsigma$ 
components below. The lengths of the vertical bars are proportional to their 
relative strength.
}
    \label{fig:PEPSI}
\end{figure}

Previous studies of Ap stars revealed that their magnetic fields cover the 
whole stellar surface and have a large-scale structure often resembling a 
single dipole whose axis is inclined with respect to the stellar rotation 
axis. Because the measured magnetic field strength depends on the viewing 
angle of the observer, that is, on the rotation phase of the star, it is 
important for our analysis that we know the strength of the mean longitudinal 
magnetic field and the mean magnetic field modulus at the time of 
observations. According to 
\citet{2021MNRAS.508L..17H}, 
the mean longitudinal magnetic field was gradually decreasing from $-$922\,G 
in 2011 to $-$572\,G in 2020, and the mean magnetic field modulus was 
gradually decreasing from 3.80\,kG in 2011 to 3.58\,kG in 2017. Our most recent 
observations\footnote{\url{https://doi.org/10.17876/data/2024_1}} 
of $\upgamma$\,Equ with the Potsdam Echelle Polarimetric and Spectroscopic 
Instrument 
\citep[PEPSI;][]{2015AN....336..324S}
installed at the $2\times8.4$\,m Large Binocular Telescope (LBT) were 
obtained on 2022 September 15, just thirteen days before the CRIRES$+$ 
observations. They show an even lower field modulus of 3.36\,kG, confirming the 
decreasing trend for the field strength. In Fig.~\ref{fig:PEPSI}, we present 
magnetically split components of the Zeeman doublet \ion{Fe}{ii} at 
6149.25\,\AA{} used for the measurements of the mean magnetic field modulus
in different years using different instruments. The PEPSI observations were
made using the cross-dispersers CD\,II, covering $4236-4770$\,\AA,{} and 
CD\,IV, covering $5391-6289$\,\AA{} with a spectral resolution of 
$R \sim 130\,000$ corresponding to 0.06\,\AA{} at 7600\,\AA.  For more 
details on observations with PEPSI and data reduction, we refer the reader to
\citet{2023A&A...674A.118S} 
and 
\citet{2021MNRAS.508L..17H}.

Since mean longitudinal magnetic field measurements of $\upgamma$\,Equ after 
2020 are not available in the literature, we relied on the work of\
\citet{2018AstBu..73..463S}, for example, 
who discussed the variability of the longitudinal field phase curve using 
a periodogram analysis. According to this work, the strength of the field had 
to be in the range of $-$200 -- 
$-$300\,G at the time of the CRIRES$^{+}$ observations.

\begin{figure}
    \centering
\includegraphics[width=1.0\columnwidth]{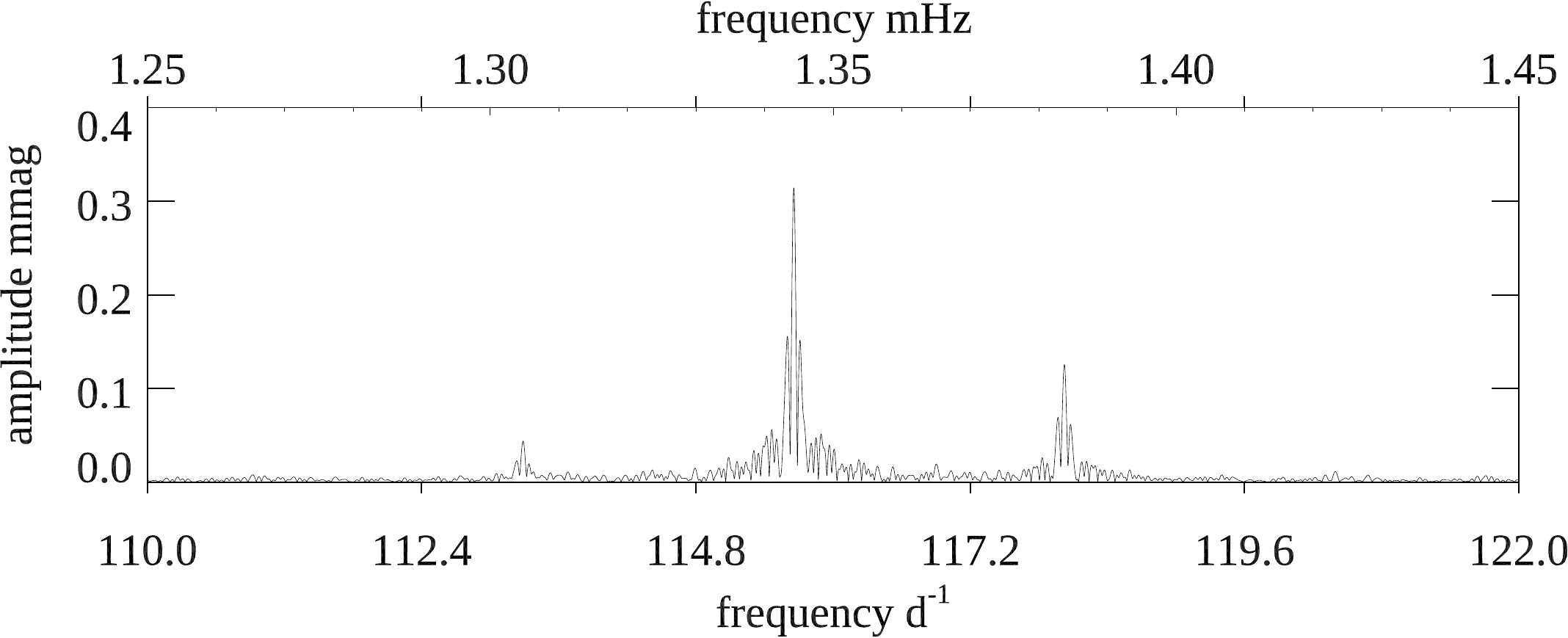} 
    \caption{
Frequency spectrum of $\upgamma$\, Equ based on Sector 55 TESS photometry.
}
  \label{fig:TESS}
\end{figure}

To be able to study the magnetic field variability over the pulsational 
cycle, we need to assign the pulsational phase to each of our spectra. 
Because $\upgamma$\,Equ is known to have multiple pulsation periods near 
12\,min
\citep{1983MNRAS.202....1K,1988ApJ...330L..51L, 1996MNRAS.282..243M,2008A&A...480..223G},
we used more recent Transiting Exoplanet Survey Satellite
\citep[\emph{TESS};][]{2015JATIS...1a4003R}
observations to identify the frequency with the highest amplitude at a time 
closest to the spectroscopic observations.

$\upgamma$\,Equ was observed by \emph{TESS} in Sector 55 from 2022 August 5 
to 2022 September 1. The analysis of these data is illustrated in 
Fig.~\ref{fig:TESS}, showing three main frequencies. The highest peak with 
an amplitude of 0.315\,mmag is at $115.5654 \pm 0.0001$\,d$^{-1}$ (1.3386\,mHz; 
or $P = 747.0404\,{\rm s} = 12.45067$\,min). The other two peaks with 
amplitudes of 0.127\,mmag and 0.042\,mmag are at 118.0236\,d$^{-1}$ 
(1.3660\,mHz) and 113.2873\,d$^{-1}$ (1.3112\,mHz), respectively.

These three frequencies are in complete agreement with the frequencies 
derived independently by 
\citet{Daniel2024}.
As those authors noted, we also see the harmonic of the highest amplitude 
peak and one combination frequency. However, it is notable that the 
frequencies previously derived for  $\upgamma$\,Equ in other studies 
\citep{1983MNRAS.202....1K,1988ApJ...330L..51L, 1996MNRAS.282..243M,2008A&A...480..223G} 
range from $113 - 123$\,d$^{-1}$ and in general do not coincide with the 
frequencies derived from the TESS data. Possible explanations are that the 
star shows mode changes, that different modes are detected at different 
atmospheric depths (since the studies were both spectroscopic and 
photometric with different filters), or that there have been problems with 
aliasing in the ground-based data, which present significant time gaps.  
Importantly, the frequencies derived from the  TESS data do not suffer 
significant aliasing, and the pulsation amplitude is stable over the 27-d 
time span of the Sector~55 data. Since the spectroscopic data were taken only a 
month after the TESS data, we assume that this mode stability continued over 
this short time span. Because of the possibility of some pulsation amplitude 
change, however, we cannot directly relate the photometric pulsation 
amplitude to the amplitude of the magnetic variations over the pulsation 
cycle. That requires simultaneous observations. 

The frequency separations of the three peaks seen in Fig.~\ref{fig:TESS} 
are 27\,$\upmu$Hz, which is plausibly half the large separation; hence, we 
suggest that the observed modes are $\ell = 2, 1, 2$ alternating-degree 
modes. We note that the modulation timescale for the outer mode frequencies 
to beat with the main mode frequency is about 10\,hr; hence, these 
frequencies are not resolved in the 56\,min of the CRIRES$+$ time series 
presented in this paper.


\section{Analysis of CRIRES$+$ spectroscopic time series}\label{sec:analysis}

\begin{figure}
    \centering
    \includegraphics[width=0.90\columnwidth]{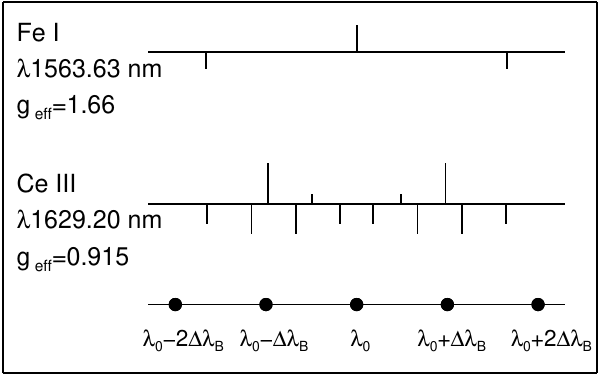}
    \caption{
Zeeman patterns of \ion{Fe}{i} triplet and \ion{Ce}{iii} pseudo-doublet with 
corresponding Land\'e factors $g_{\rm eff}$. Each Zeeman component is 
represented by a vertical bar whose length is proportional to its relative 
strength. The $\uppi$ components are presented above the horizontal 
wavelength axis, and the $\upsigma_\pm$ components are shown below it. The 
unit length of the wavelength axis is $\Delta\lambda_B,$ defined as the 
wavelength shift from the line centre of the $\upsigma$ components in a 
normal Zeeman triplet with a Land\'e factor of $g_{\rm eff}$=1.
}
    \label{fig:patterns}
\end{figure}

In the recorded CRIRES$+$ spectroscopic time series, two almost blend-free 
magnetically split lines, the Zeeman triplet \ion{Fe}{i} at 1563.63\,nm and 
the pseudo-doublet line \ion{Ce}{iii} at 1629.20\,nm, appear to be most 
suitable to study the pulsational variability of the mean magnetic field 
modulus. Their identification is based on the work of 
\citet{2019ApJ...873L...5C}.
Zeeman patterns for both lines and the corresponding effective Land\'e 
factors are presented in Fig.~\ref{fig:patterns}. 

\begin{figure}
   \centering
   \includegraphics[width=0.90\columnwidth]{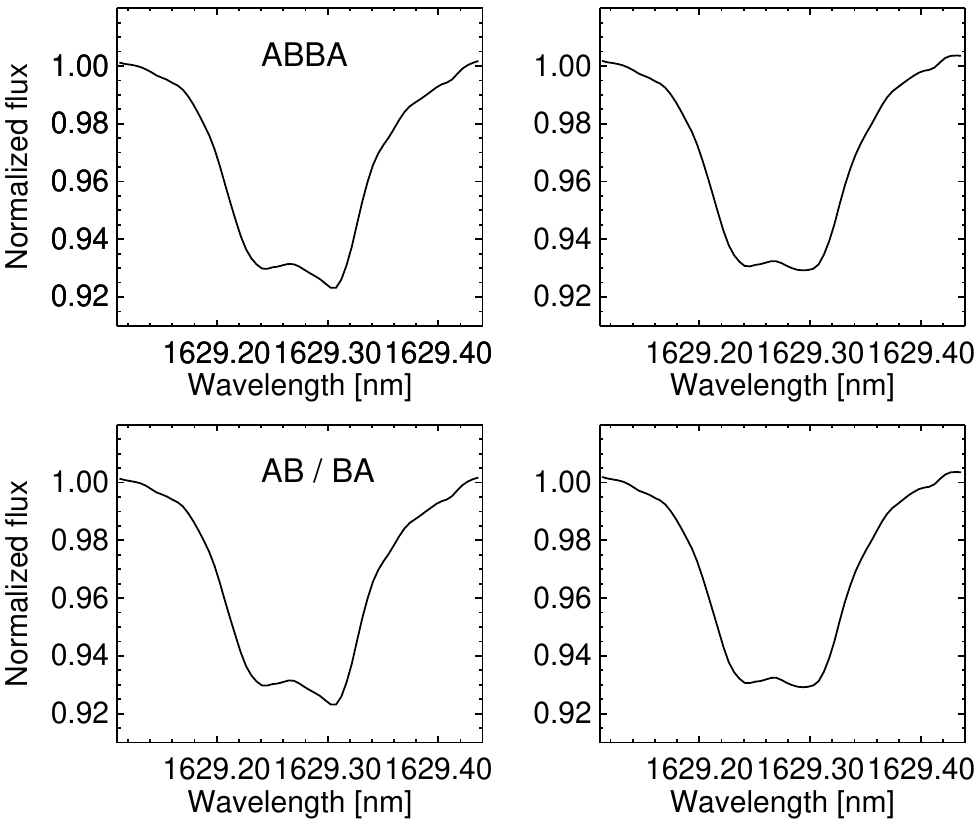}
    \caption{
Mean \ion{Ce}{iii}\,1629.2\,nm line profiles in their original shape (left) 
and corrected  for telluric contributions (right), calculated for both 
sequences ABBA and AB/BA.
}
    \label{fig:telluric}
\end{figure}

While the spectral region around the \ion{Fe}{i}\,1563.63\,nm line is 
perfectly clean and is not affected by telluric lines, the 
\ion{Ce}{iii}\,1629.2\,nm line shows a small telluric contribution on the 
red side of the line profile. It was removed using the ESO SkyCalc based on 
the Cerro Paranal Sky Model 
\citep{2012A&A...543A..92N,2013A&A...560A..91J}.
The original and the mean \ion{Ce}{iii}\,1629.2\,nm  corrected for the 
telluric contribution profiles for both sequences ABBA and AB/BA are 
presented in Fig.~\ref{fig:telluric}.

\begin{figure}
    \centering
    \includegraphics[width=0.90\columnwidth]{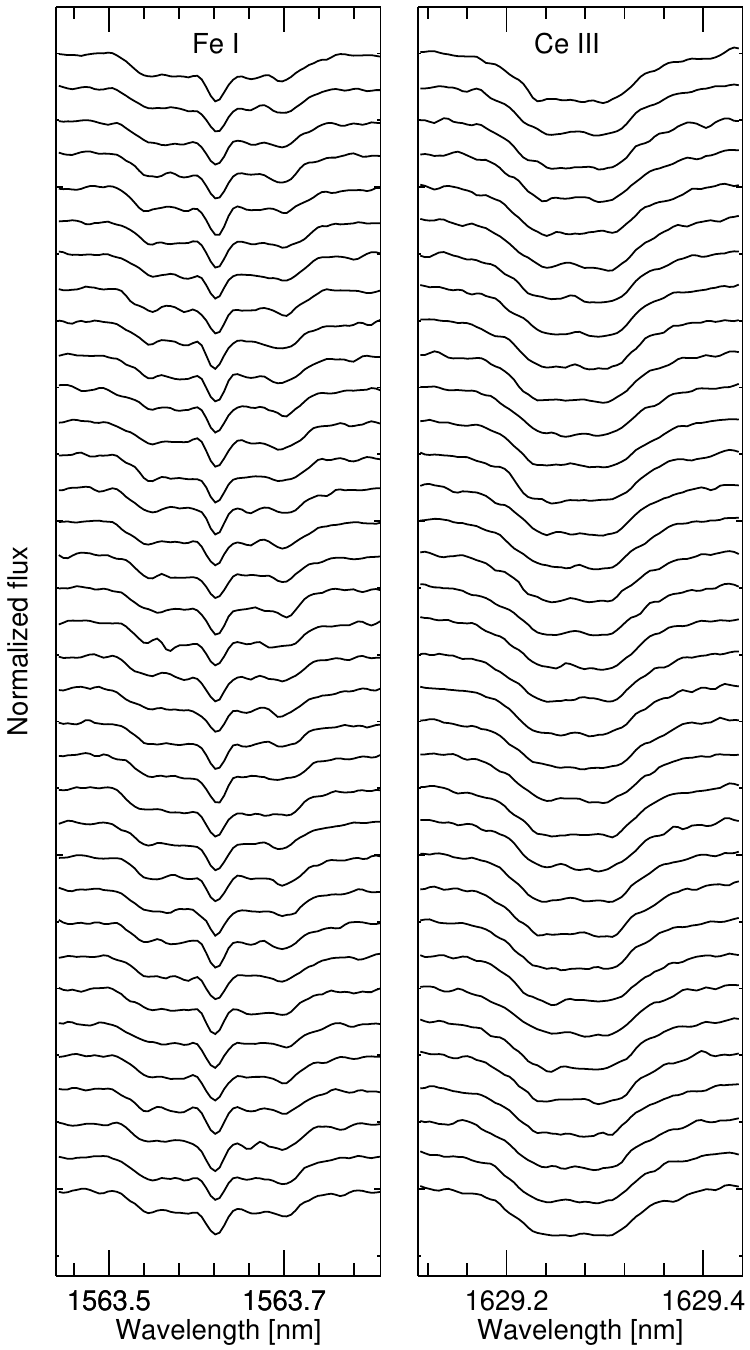}
    \caption{
Individual line profiles for \ion{Fe}{i}\,1563.63\,nm (left) and for 
\ion{Ce}{iii}\,1629.2\,nm (right) after telluric correction. The time 
increases from bottom to top. 
}
    \label{fig:allprofs}
\end{figure}

\begin{figure}
    \centering
    \includegraphics[width=0.90\columnwidth]{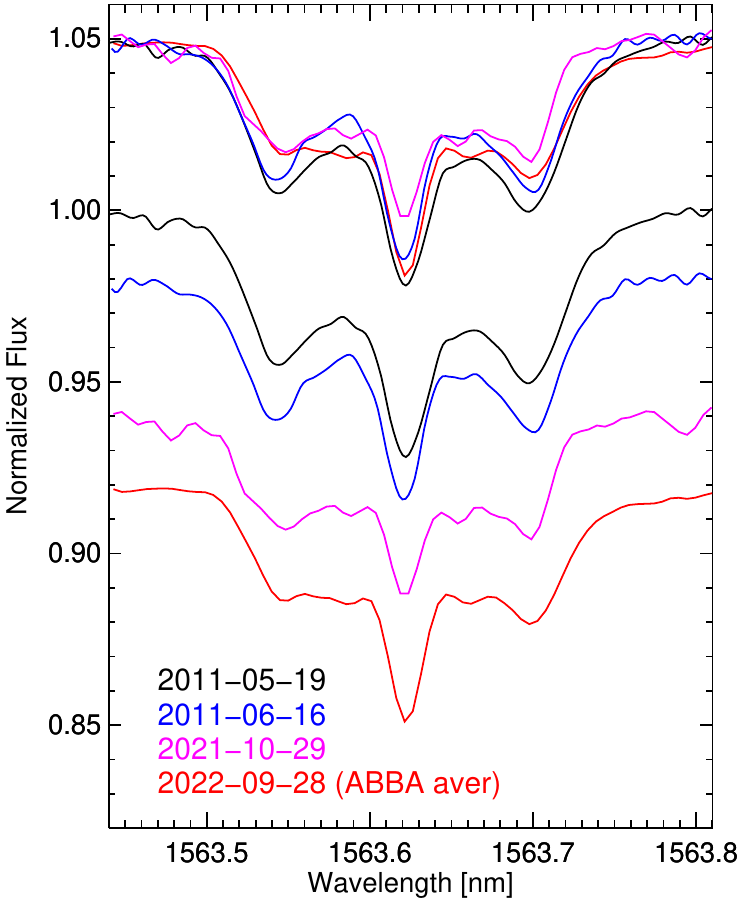}
    \caption{
Comparison of line profiles of Zeeman triplet \ion{Fe}{i}\,1563.63\,nm
between 2011 May, 2011 June, 2021 October, and 2022 September. The two 
spectra from 2011 were recorded with CRIRES, whereas for the observations in 
2021 and 2022 CRIRES$+$ was used. The overplotted profiles are presented at 
the top. 
}
    \label{fig:oldnew}
\end{figure}

\begin{figure}
    \centering
    \includegraphics[width=0.38\textwidth]{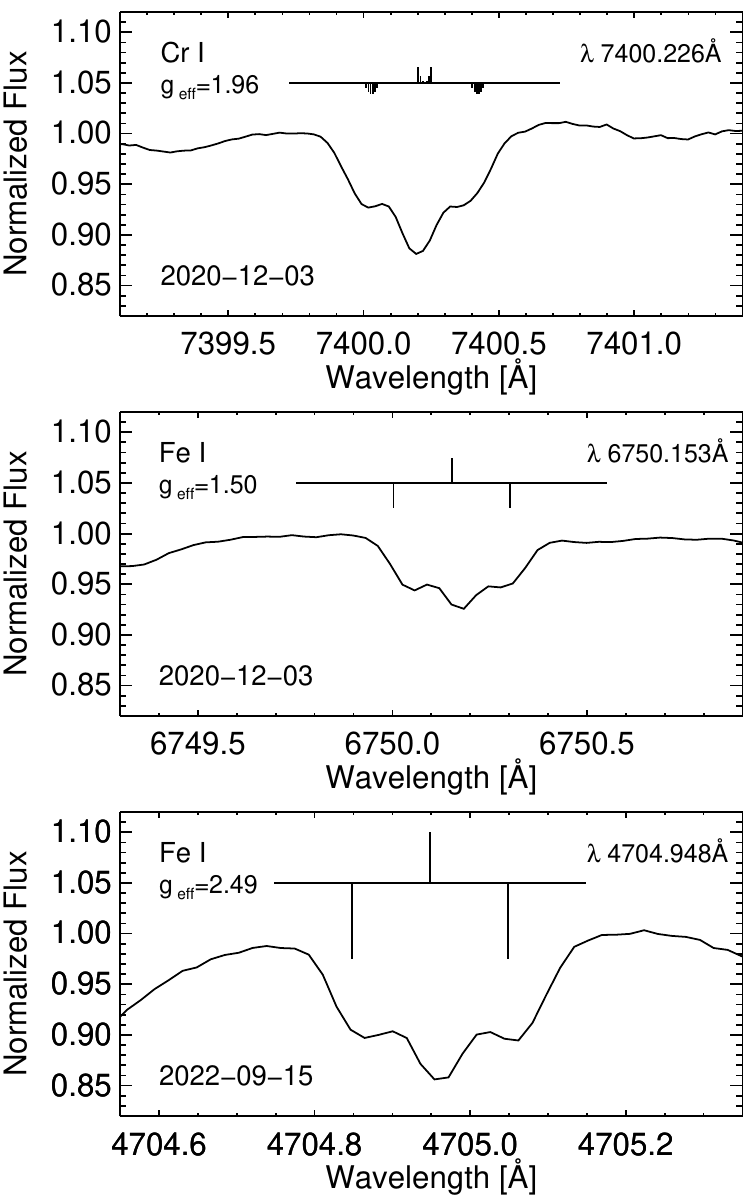}
    \caption{
Line profiles of two Zeeman triplets and one pseudo-triplet observed with 
PEPSI in the visual spectral region in 2020 and 2022. The shape of the line 
profiles and the separation between the $\uppi$ and $\upsigma$ components are 
completely different compared to the appearance of the Zeeman triplet 
\ion{Fe}{i}\,1563.63\,nm observed on 2022 September 28 with CRIRES$+$. 
Zeeman patterns for the spectral lines are presented at the top of each panel. 
}
    \label{fig:visual}
\end{figure}

Individual line profiles' time series obtained over 56\,min using the ABBA 
sequence are presented in Fig.~\ref{fig:allprofs}. Surprisingly, in contrast 
to the high-resolution CRIRES 
\citep{2003SPIE.4843..223K} 
and CRIRES$+$ observations of the \ion{Fe}{i}\,1563.63\,nm line obtained in 
previous years, the line profile shape recorded in our observations in 2022 
shows a significant change, displaying rather flat $\upsigma$ components and 
not very clear separation between them and the $\uppi$ component. While the 
presence of the $\upsigma$ component on the red side of the line profile can 
still be perceived, the position of the blue-side $\upsigma$ component is not 
clearly defined. In Fig.~\ref{fig:oldnew} we present very different shapes 
of the line profiles of the Zeeman triplet \ion{Fe}{i}\,1563.63\,nm observed 
in different years using CRIRES and CRIRES$+$. Observations on 2011 May 19 
and 2011 June 16 were obtained with an exposure time of 450\,s. For the 
observations on 2021 Oct 29, four frames with 20\,s of exposure time each 
were obtained using one nodding cycle. It is possible that the `filling' 
between the $\uppi$ and $\upsigma$ components that appeared in 2021--2022, 
while not present in 2011, is due to blending with a line of another element 
whose wavelength coincides with that of the \ion{Fe}{i}\,1563.63\,nm line. 
The $\sim$10 years elapsed from 2011 to 2021 may represent about one tenth of 
the rotation period, which may be enough for an overabundance spot of some 
element to have appeared on the visible stellar hemisphere as a result of 
stellar rotation. However, the measured equivalent width of this line is the 
same in 2021–-2022 as it was in 2011.

As discussed in Sect.~\ref{sec:obs}, the magnetic field strength in 
$\upgamma$\,Equ is gradually decreasing and will probably approach the 
rotation phase of the best visibility of the magnetic equator in a few 
years, where the longitudinal magnetic field component is expected to be 
weaker and the transverse field component becomes the strongest. For a 
transverse field, the $\uppi$ components become, to the first order, twice 
as strong as the $\upsigma_{+}$ and the $\upsigma_{-}$ components. 
Interestingly, in contrast with the observations in the near-IR, the 
profiles of the Zeeman triplet lines recorded by us in the visual wavelength 
region in 2020 and 2022 using PEPSI show much clearer $\upsigma$ components, 
which can be more easily used for the measurement of the magnetic field 
modulus. The different appearance of the triplets can probably be explained 
by different iron vertical stratification between the visual and near-IR 
regions. There may also be a significant difference in the optical depth 
from which the continuum is issued between the visible and the IR, and, as 
a consequence, there could be different (de)saturations of the split line 
components between the two wavelength regions. This alone may potentially 
result in differently looking triplets. In Fig.~\ref{fig:visual}, we 
present line profiles of two Zeeman triplets and one Zeeman pseudo-triplet 
observed with PEPSI in the visual spectral region.

\begin{figure}
    \centering
    \includegraphics[width=0.90\columnwidth]{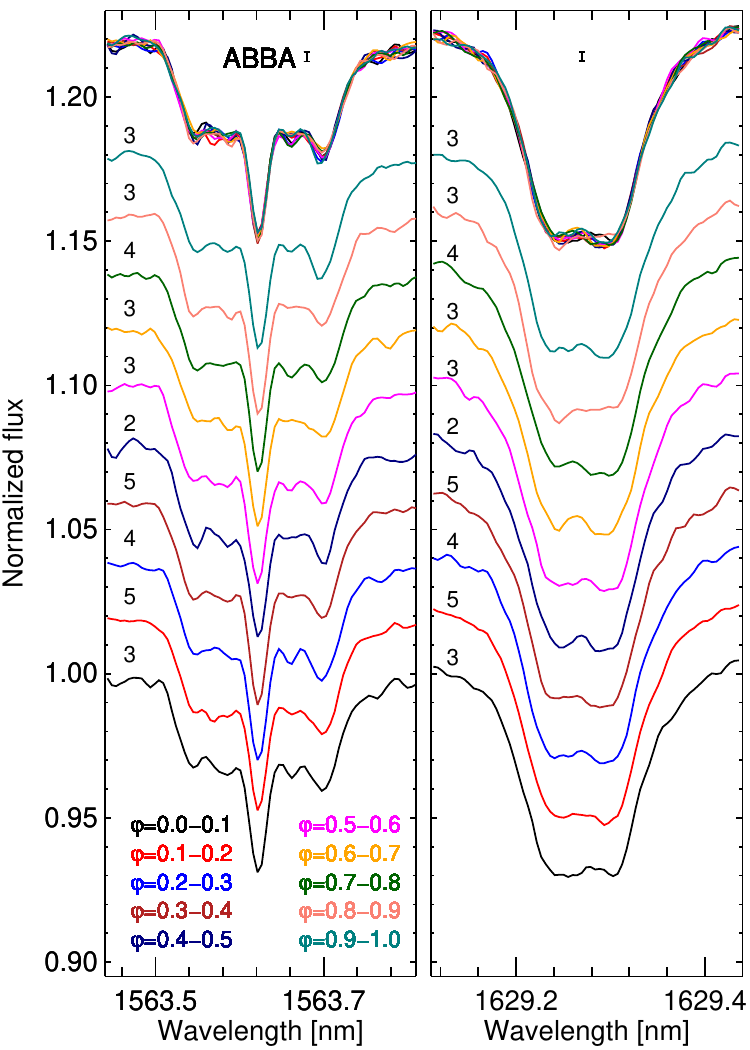}
    \caption{
Phase-binned mean line profiles of \ion{Fe}{i} line (left) and \ion{Ce}{iii} 
line (right) calculated using the ABBA sequence. For each bin, profiles are 
plotted with different colours and shifted upward for better visibility. The 
number of spectra within each bin is indicated close to each profile. At the 
top, we show the overplotted profiles and the corresponding error bars.
}
    \label{fig:meanphaprof01}
\end{figure}

As shown in Fig.~\ref{fig:allprofs}, several absorption bumps appear from 
time to time not only on both sides of the $\uppi$ component in the line 
profile of \ion{Fe}{i}\,1563.63\,nm, but also in the core of the 
\ion{Ce}{iii}\,1629.2\,nm line. Since the $S/N$ of the individual spectrum is 
rather low, it is not clear whether the observed variability is intrinsic or 
is caused by the noise. To investigate the nature of these bumps in more 
detail, we determined, for each individual spectrum calculated in the ABBA 
sequence, the corresponding pulsation phase assuming the pulsational period 
of 12.45067\,min as determined from the TESS observations. To increase the 
$S/N$, the spectra have been binned with the phase bin of 0.1. The number of 
individual spectra in each bin varies from two to five, allowing us to 
increase the $S/N$ by a factor of 1.4 to 2.2, respectively. The plots for 
the mean spectra obtained within each bin are displayed in 
Fig.~\ref{fig:meanphaprof01} together with the information on the number of 
spectra included in each bin. While no radial velocity variability is 
detected in the overplotted line profiles, we observe that the depth of the 
$\uppi$ component of the \ion{Fe}{i}\,1563.63\,nm line is slightly variable; 
moreover, the fractions of the profile around the $\uppi$ component show 
variable absorption bumps. A comparison of the variability of the profiles 
that have the largest numbers of individual spectra per bin and therefore a 
higher $S/N$ (phases 0.0--0.1, 0.1--0.2, 0.2--0.3, 0.3--0.4, 0.7--0.8) shows 
that absorption bumps become stronger in the red part of the profile in the 
phase bins 0.0--0.1, 0.2--0.3, and 0.7--0.8. In the case of the 
\ion{Ce}{iii} line, the separation between the split components becomes much 
more clear in several phase bins, the clearest case being the phase bin 
0.4--0.5.

\begin{figure*}
    \centering
    \includegraphics[width=0.4\textwidth]{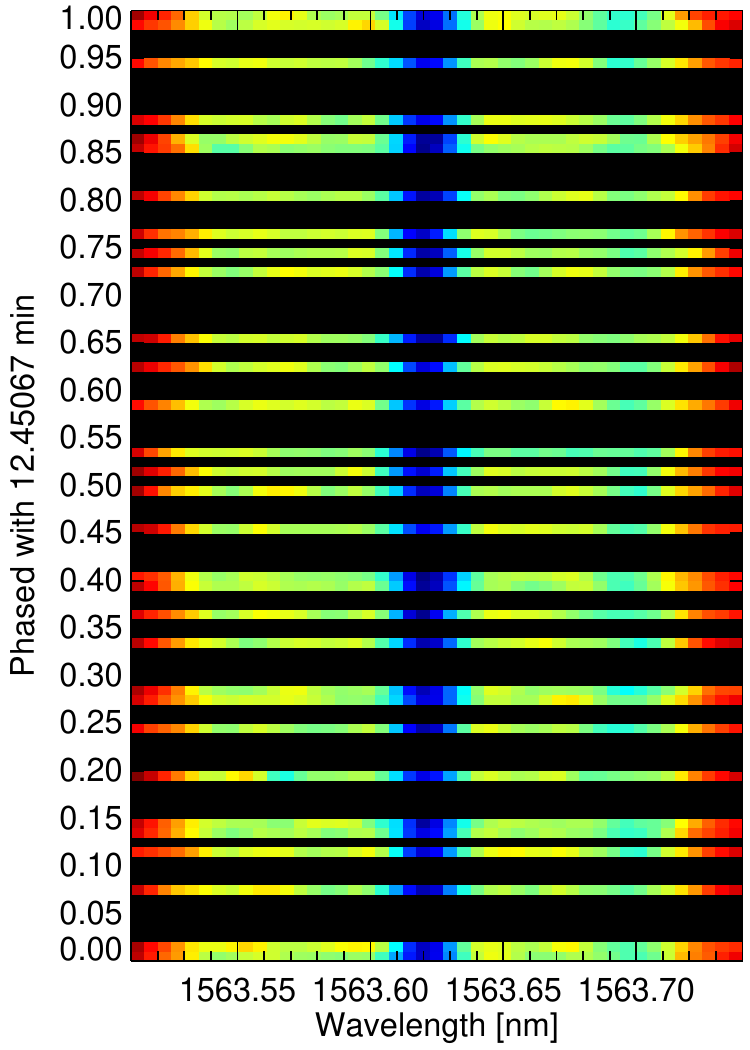}
    \includegraphics[width=0.4\textwidth]{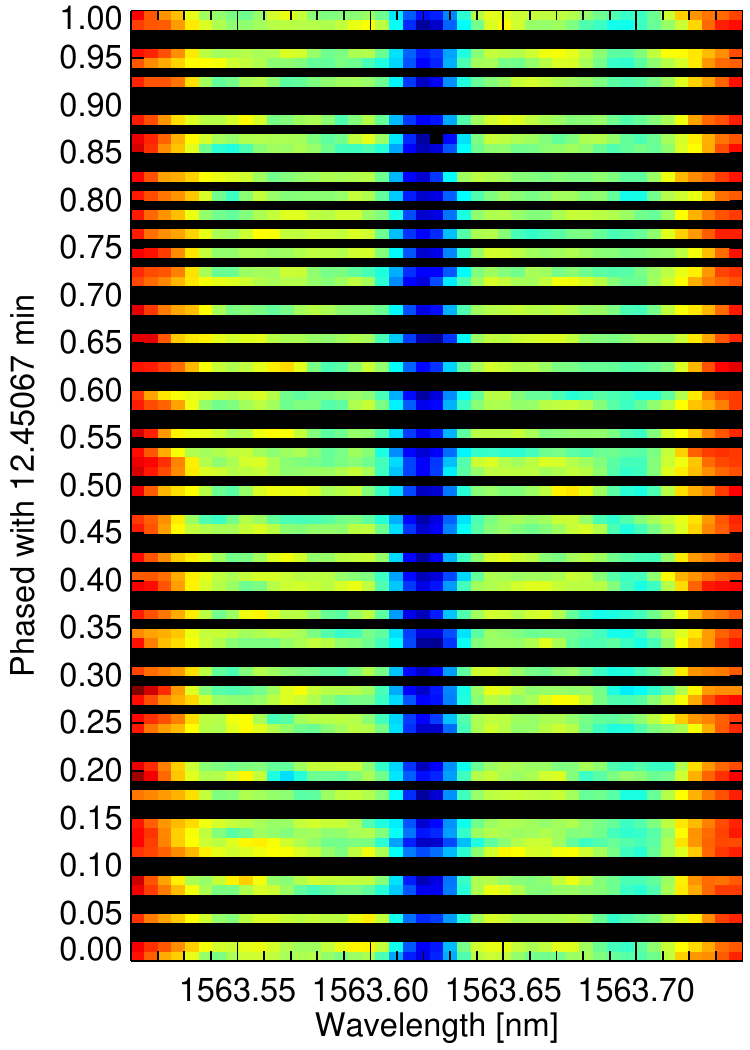}
    \caption{
Dynamical spectrum of whole \ion{Fe}{i}\,1563.63\,nm line. The bluest 
colour is for the deepest absorption at the $\uppi$ component, whereas red 
is at the continuum level. 
\emph{Left:} Profiles reduced using the ABBA sequence.
\emph{Right:} Profiles reduced using the AB/BA sequence.
}
    \label{fig:profdyn1563}
\end{figure*}

\begin{figure*}
    \centering
    \includegraphics[width=0.4\textwidth]{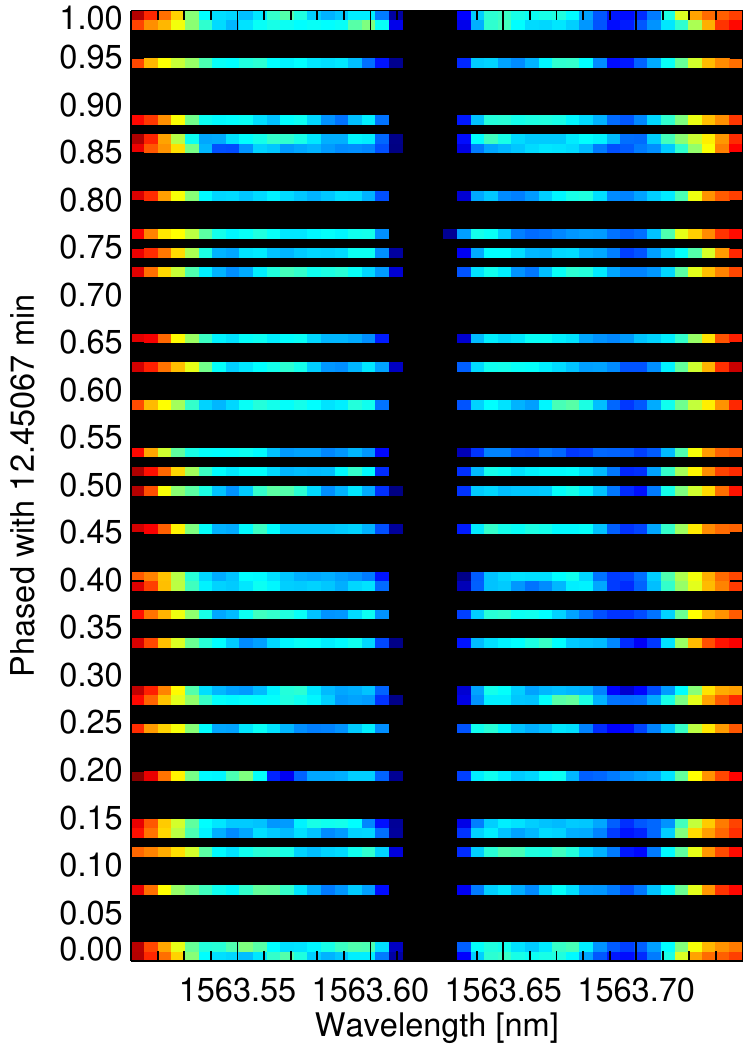}
    \includegraphics[width=0.4\textwidth]{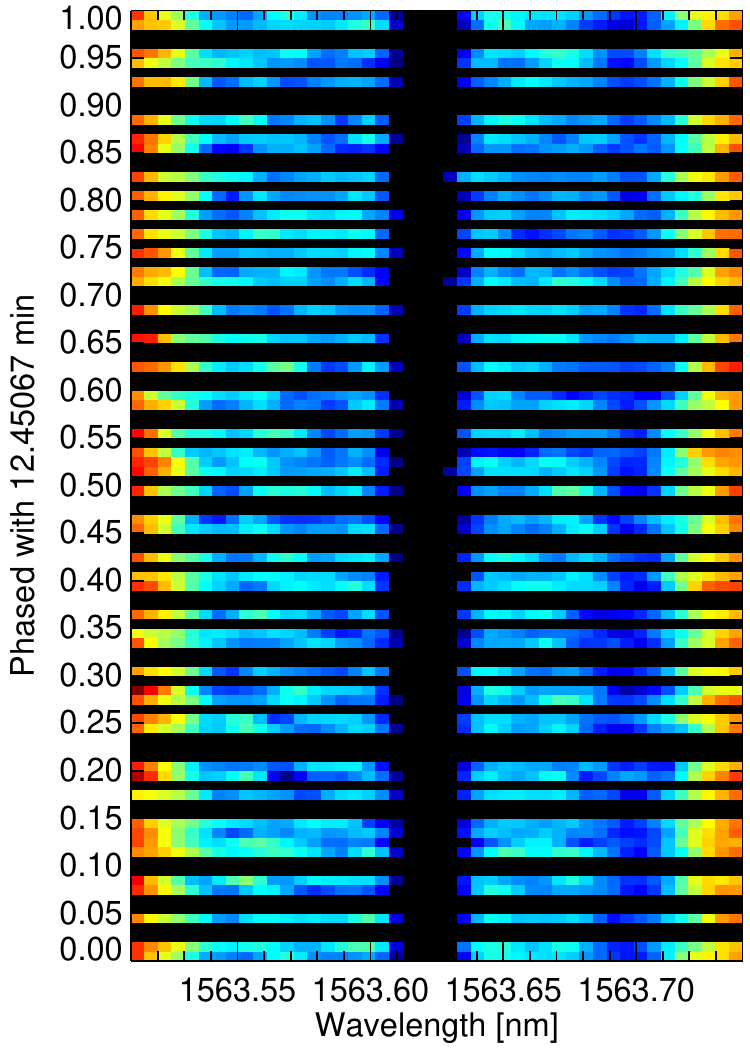}
    \caption{
As Fig.~\ref{fig:profdyn1563}, but $\uppi$ component has been masked out in 
order to show the $\upsigma$ components better.
\emph{Left:} Profiles reduced using ABBA sequence.
\emph{Right:} Profiles reduced using AB/BA sequence.
}
    \label{fig:allnopidyn1563}
\end{figure*}

\begin{figure*}
    \centering
    \includegraphics[width=0.4\textwidth]{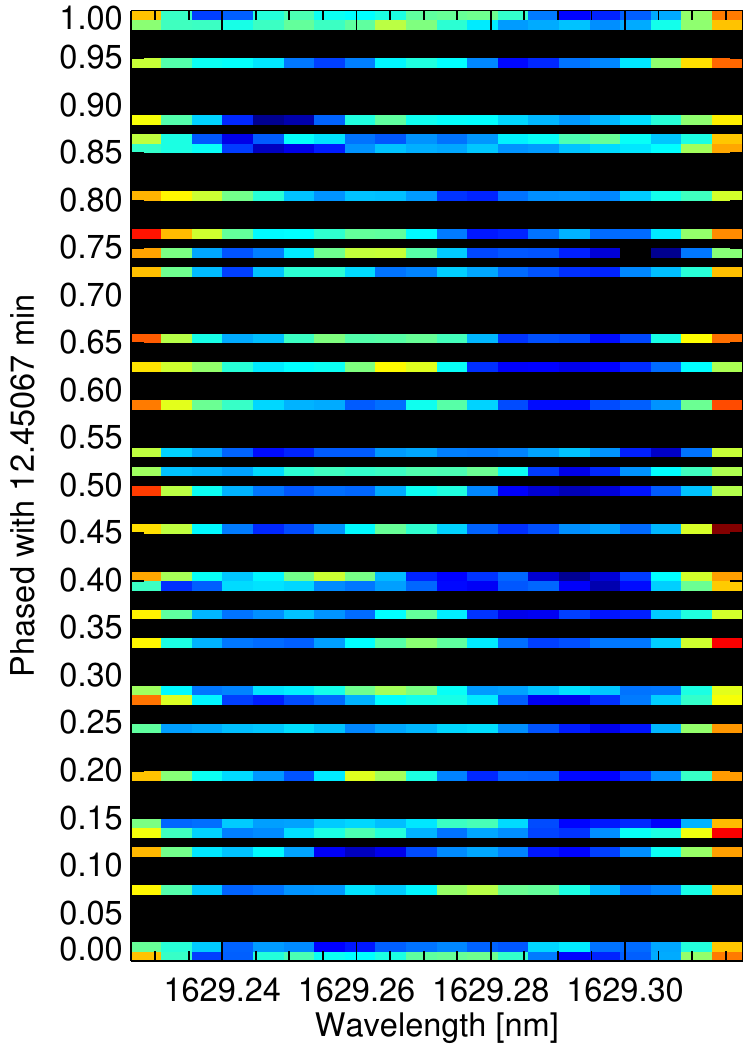}
    \includegraphics[width=0.4\textwidth]{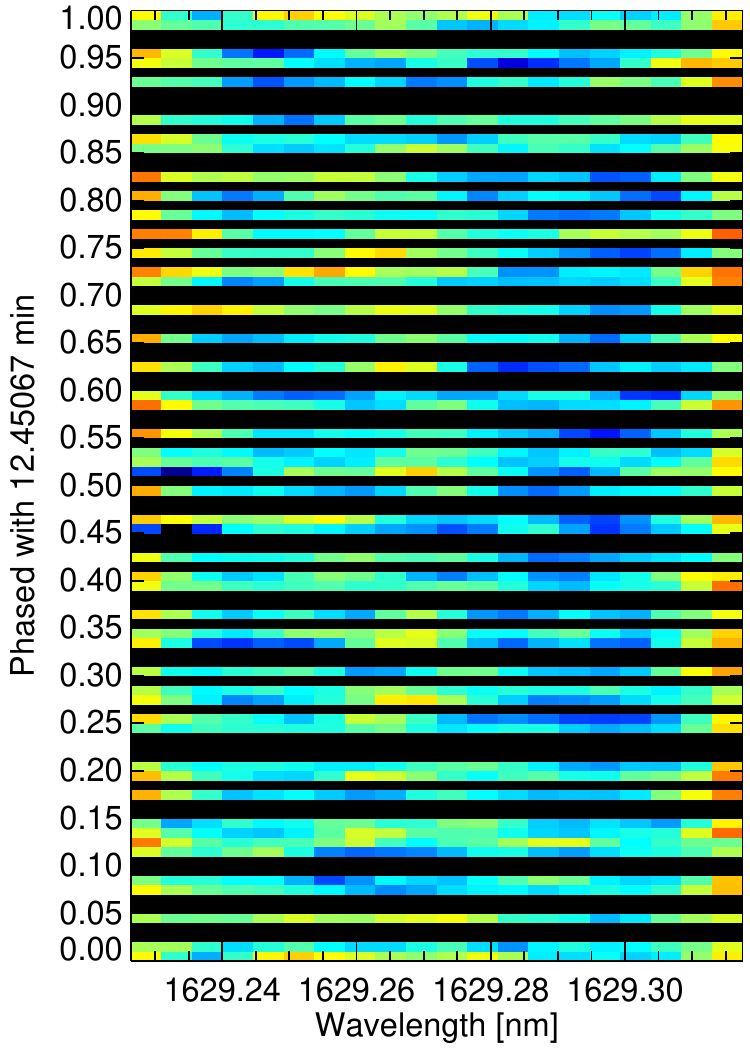}
    \caption{
Dynamical spectrum of line core of \ion{Ce}{iii}\,1629.2\,nm. The bluest 
colour is for the deepest absorption, whereas red is for the shallower 
absorption. Black is used for phases not covered by the observations.
\emph{Left:} Profiles reduced using ABBA sequence.
\emph{Right:} Profiles reduced using AB/BA sequence.
}
    \label{fig:coredyn1629}
\end{figure*}

To delve deeper into the character of the line profile variability, in 
Figs.~\ref{fig:profdyn1563}--\ref{fig:coredyn1629} we present dynamical 
spectra for the observed profiles of \ion{Fe}{i}\,1563.63\,nm and the 
\ion{Ce}{iii}\,1629.2\,nm lines phased with the pulsation period and using 
both the ABBA and the AB/BA sequences. The blue-green colour in  
Fig.~\ref{fig:profdyn1563} constructed for the \ion{Fe}{i}\,1563.63\,nm line 
clearly shows the variability of the $\upsigma$ component on the red side; it 
changes intensity and becomes significantly broader in some phases. The 
$\upsigma$ component on the blue side almost disappears in some phases, that 
is, between the phases 0.0--0.3. This behaviour is confirmed in the dynamical 
spectra presented in Fig.~\ref{fig:allnopidyn1563}, where we mask out the 
$\uppi$ component to achieve a better contrast. Interestingly, absorption 
bumps discovered in the flat blue part of the profile appear shifted and 
stronger close to the pulsation phase 0.20, whereas the absorption bump on 
the red side becomes strongest in the pulsation phase around 0.7--0.8. In 
the plot corresponding to the AB/BA sequence, a sort of arc appears in 
the phases 0.80--0.90. The inspection of the dynamical spectra in 
Fig.~\ref{fig:coredyn1629} calculated for the \ion{Ce}{iii}\,1629.2\,nm line
shows that this line is also variable where the depth of the right component 
of the split profile almost disappears in the pulsation phases between 
0.25--0.30 and 0.85--0.90.


\section{Discussion}

\begin{figure}
    \centering
    \includegraphics[width=0.9\columnwidth]{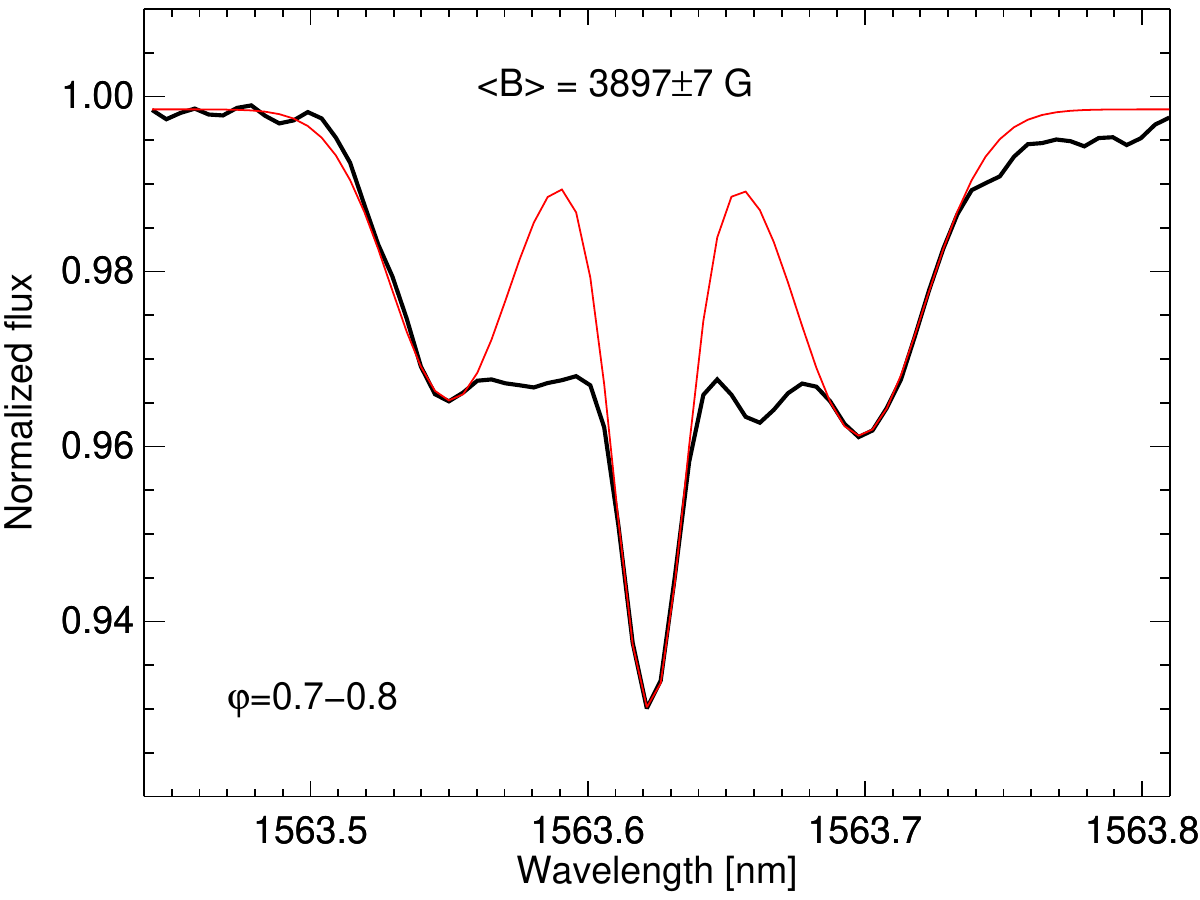}
    \includegraphics[width=0.9\columnwidth]{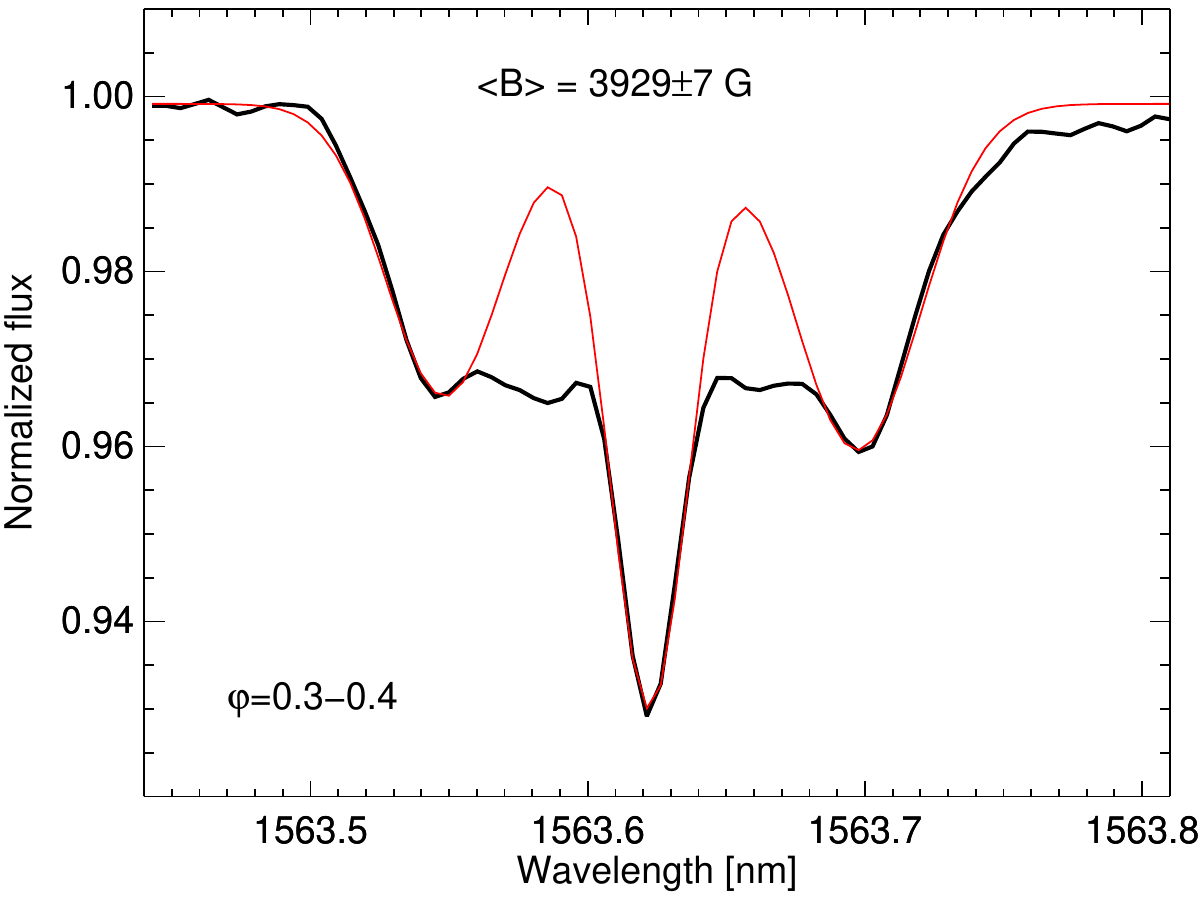}
    \caption{
Zeeman triplet \ion{Fe}{i}\,1563.63\,nm observed in phases 0.3--0.4 and 
0.7--0.8. The black lines present the observed profiles and the red lines 
are used to show the triple-Gaussian fits.
}
    \label{fig:fesplit}
\end{figure}

\begin{figure}
    \centering
    \includegraphics[width=0.9\columnwidth]{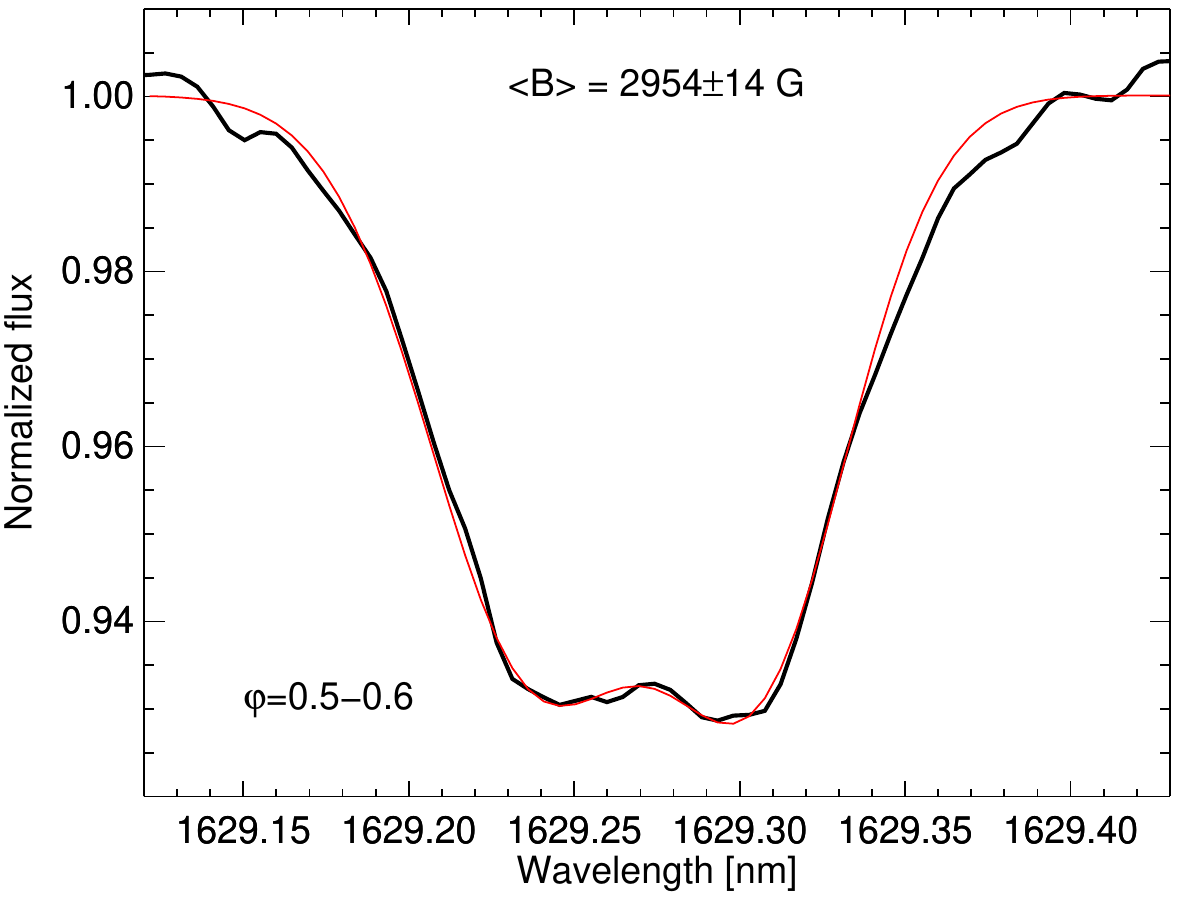}
    \includegraphics[width=0.9\columnwidth]{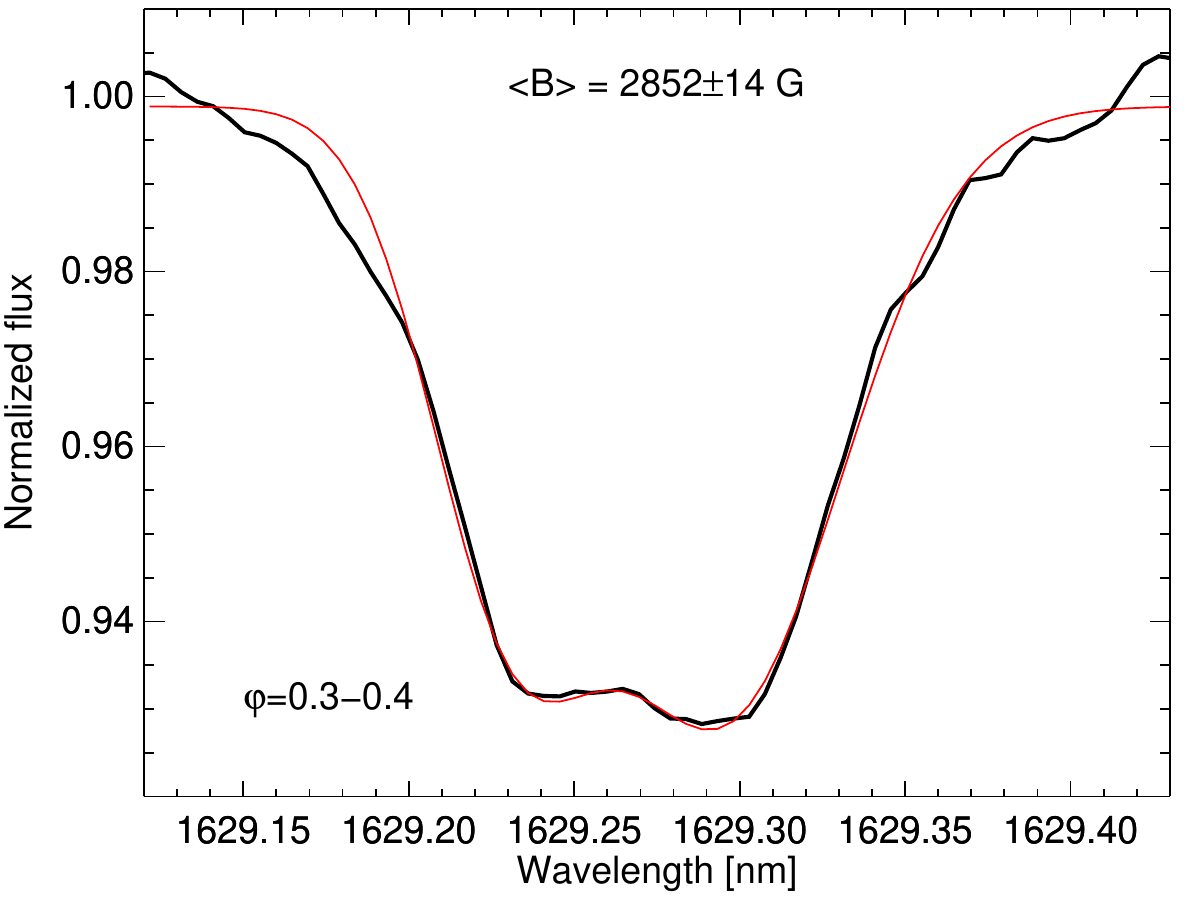}
    \caption{
Zeeman pseudo-doublet \ion{Ce}{iii}\,1629.2\,nm observed in phases 0.3--0.4 
and 0.5--0.6. The black lines present the observed profiles and the red 
lines are used to show the double-Gaussian fits.
}
    \label{fig:cesplit}
\end{figure}

Our study of the pulsation behaviour of the magnetic roAp star $\upgamma$\,Equ 
in the near-IR using CRIRES$+$ time series over 56\,min reveals a distinct 
variability of the magnetically split lines \ion{Fe}{i}\,1563.63\,nm and 
\ion{Ce}{iii}\,1629.2\,nm. The character of this variability appears, however, 
rather complex, probably due to a significant decrease in the strength of the 
longitudinal field component and an increase in the strength of the 
transverse field component over the last decade. The reason for the observed 
absorption bumps around the $\uppi$ component in the Zeeman triplet 
\ion{Fe}{i}\,1563.63\,nm is not clear. We simply speculate that they are 
possibly related to the near-surface convection in the presence of the 
stratification of iron.

In contrast to the older high-resolution CRIRES 
\citep{2003SPIE.4843..223K} 
observations, the line profile shape of the Zeeman triplet 
\ion{Fe}{i}\,1563.63\,nm  recorded in our observations in 2022 displays 
rather flat $\upsigma$ components with unclear separation between them 
and the $\uppi$ component. Assuming that the outer absorption dips on both 
sides of the $\uppi$ component in the line profile of the Zeeman triplet 
\ion{Fe}{i}\,1563.63\,nm present the $\upsigma$ components, we tried to 
measure the mean magnetic field modulus using the line profiles recorded in 
the pulsational phase bins 0.3--0.4 and 0.7--0.8. The results of our 
measurements with a triple-Gaussian fit are presented in 
Fig.~\ref{fig:fesplit}. Interestingly, the measured modulus strengths of 
3.929\,kG in the phase bin 0.3--0.4 and 3.897\,kG in the phase bin 
0.7--0.8 are not identical, implying that the mean magnetic field modulus 
probably varies over the pulsation period with an amplitude of $32\pm7$\,G. 
On the other hand, the much stronger magnetic field modulus measured 
in the optical region contradicts the one reported for the IR (see 
Sect.~\ref{sec:obs} and Fig.~\ref{fig:PEPSI}). We measured a much lower 
field modulus of 3.36\,kG on 2022 September~15 using the Zeeman doublet 
\ion{Fe}{ii} at 6149.25\,\AA{}. Such a difference is not unexpected, for 
various reasons. First, the mean magnetic field modulus is the average of 
the field modulus over the visible stellar hemisphere, averaged by the local 
emergent line intensity. Because the limb darkening is not the same in the 
visible and in the IR, the line intensity distribution over the stellar disk 
is different between the two wavelength regions, which leads to a different 
weighting of the contributions of the various parts of the stellar surface 
in the magnetic field averaging process. Furthermore, the continuum optical 
depth of the line-forming layer is not the same in the IR as in the visible. 
Accordingly, the saturation degree of the individual $\uppi$ and $\upsigma$ 
components of lines with anomalous Zeeman patterns may be different, which 
affects the interpretation of the observed line profiles in terms of 
magnetic field strength. Finally, while vertical gradients of the magnetic 
field strength may not be required to justify the difference in the field 
modulus values derived from the consideration of diagnostic lines in 
different wavelength ranges, their existence and their potential effect on 
the magnetic field measurements are plausible.  

Much lower magnetic field modulus values are determined using the Zeeman 
pseudo-doublet \ion{Ce}{iii}\,1629.2\,nm. In the 0.3--0.4 phase bin, we 
measure 2.954\,kG, whereas the value 2.852\,kG is determined for the spectra 
in the 0.5--0.6 phase bin. These results presented in Fig.~\ref{fig:cesplit} 
imply that the mean magnetic field modulus probably varies over the 
pulsation period with the amplitude of $102\pm14$\,G. Since iron is reported 
to be gravitationally settled in the stellar atmospheres of roAp stars 
compared to the rare earth elements that are usually concentrated in the 
upper atmospheric layers 
\citep[e.g.][]{2013A&A...552A..28N},
it is expected that the lines of different elements trace the magnetic field 
strength and the pulsation amplitudes differently. However, because of the 
extremely abnormal chemical composition of the atmospheres of magnetic roAp 
stars, the calculation of the vertical element abundance distribution is a 
very complex process requiring the fitting of spectroscopic, photometric, 
and magnetic data using realistic self-consistent atmospheric models. There 
is also a number of previous studies indicating the possible presence of 
vertical magnetic field gradients in roAp stars 
\citep[e.g.][]{2004A&A...422L..51N,2018MNRAS.477.3791H},
with the mean magnetic field and longitudinal field strengths changing with 
atmospheric depth. The significant increase of the magnetic field in the 
deeper atmospheric layers would explain the difference in our measurements 
of the magnetic field modulus using the iron Zeeman triplet and the Ce 
Zeeman pseudo-doublet. 

Even though the combination of a vertical gradient of the magnetic field and 
of stratification can plausibly contribute to the observed difference between 
the magnetic field strengths derived from consideration of the \ion{Fe}{i} 
and \ion{Ce}{iii} lines, there is also another effect that must definitely 
play a role. Because the Zeeman pattern of the \ion{Fe}{i} line at 1563.63\,nm 
is a pure triplet, the determination of the mean magnetic field modulus 
from the separation of the split components is almost approximation-free. In 
particular, it does not depend on radiative transfer effects. By contrast, 
to determine $\langle B \rangle$ from the splitting of the line 
\ion{Ce}{iii}\,1629.2\,nm, whose Zeeman pattern is a pseudo-doublet, one 
must use the weak-line approximation. By doing so, one ignores saturation 
effects. The outer $\uppi$ components have a much higher relative strength 
than their inner counterparts. Therefore, the outer $\uppi$ components will 
saturate first. As a result of radiative transfer, the ratio of the depths 
of the inner-to-outer $\uppi$ components will become greater than the ratio 
of their strengths. As a result, the wavelength separation of the blue and 
red line components will be less than it would be for the \ion{Fe}{i} line, 
for which radiative transfer effects are negligible. Thus, a smaller value 
of the mean magnetic field modulus will be derived. The effect is similar 
to what we observe. One way to achieve a better assessment of the 
importance of this effect (which definitely must be present) would be to 
observe a diagnostic line with a pseudo-doublet (or pseudo-triplet) pattern 
for which the radiative transfer effects should lead to an increase in the 
derived value of $\langle B \rangle$. Unfortunately, no suitable line in the 
wavelength range covered by our spectra was observed.

On the other hand, another effect that may also lead to differences in the 
values of $\langle B \rangle$ derived from lines of different ions are
horizontal inhomogeneities, which may lead to the different sampling of the 
local magnetic field strength. Currently, we do not have enough information 
to assess the significance of this effect, but  its potential 
contribution is still worth mentioning.

The results of our observations suggest that near-IR observations present an 
important diagnostic potential in the studies of pulsations in magnetic roAp 
stars using spectral lines with different formation  heights. Future near-IR 
studies of roAp stars with strong mean longitudinal magnetic fields 
(observed close to the best visibility of their magnetic poles) in the 
near-IR using long, high-resolution spectroscopic time series and combined 
with detailed modelling will be worthwhile in the improvement of our 
understanding of the concerned physics.

\begin{acknowledgements}

We thank the anonymous referee for their useful comments.
This work is based on observations made with ESO telescopes at the
La Silla Paranal Observatory under programme IDs 087.C-0124(A),
0108.D-0659(A), and 0109.D-0659(A) publicly available via ESO Archive.
This work is partially based on observations carried out with the PEPSI
spectropolarimeter. PEPSI was made possible by funding through the State of 
Brandenburg (MWFK) and the German Federal Ministry of Education and Research 
(BMBF) through their Verbundforschung grants 05AL2BA1/3 and 05A08BAC. LBT 
Corporation partners are the University of Arizona on behalf of the Arizona 
university system; Istituto Nazionale di Astrofisica, Italy; LBT 
Beteiligungsgesellschaft, Germany, representing the Max-Planck Society, the
Leibniz-Institute for Astrophysics Potsdam (AIP), and Heidelberg University; 
the Ohio State University; and the Research Corporation, on behalf of the 
University of Notre Dame, the University of Minnesota, and the University of 
Virginia.
This paper includes data collected by the TESS mission. Funding for the
TESS mission is provided by the NASA Science Mission Directorate.

\end{acknowledgements}

%
   \bibliographystyle{aa} 
   \bibliography{crires} 
%

\end{document}